\newcommand{\g}{$\gamma$}

\documentclass[final,5p,times,twocolumn]{elsarticle} % version journal

\usepackage{url}
\usepackage{lineno,hyperref}
\modulolinenumbers[5]
\usepackage{amssymb}
\usepackage{subfigure}
\usepackage{placeins}
\usepackage{mathtools}
\usepackage{amsmath}
\usepackage{gensymb}
\usepackage{svg}

\journal{Nuclear Instruments and Methods in Physics Research A}

\begin{document}
\begin{frontmatter}

\title{Imaging neutrons with a position-sensitive monolithic CLYC detector}

\author[1]{J.~Lerendegui-Marco\footnote{jorge.lerendegui@ific.uv.es}}
\author[1,2]{G.~Cisterna}
\author[1,3]{J. ~Hallam}
\author[1,4]{V.~Babiano-Su\'arez}
\author[1]{J.~Balibrea-Correa}
\author[1]{D.~Calvo}
\author[1]{I.~Ladarescu}
\author[1]{G.~de~la~Fuente}
\author[1]{B. Gameiro}
\author[4]{A. Sanchis-Molt\'o}
\author[1]{P. Torres-S\'anchez}
\author[1]{C.~Domingo-Pardo}

\affiliation[1]{organization={Instituto de Física Corpuscular, CSIC-Universitat de València},country={Spain}}
\affiliation[2]{organization={Universidad de Sevilla},country={Spain}}
\affiliation[3]{organization={University of Surrey},country={United Kingdom}}
\affiliation[4]{organization={Universitat de València},country={Spain}}

\begin{abstract}
In this work, we have developed and characterized a position-sensitive CLYC detector that acts as the neutron imaging layer and $\gamma$-ray Compton scatterer of the novel dual Gamma-ray and Neutron Vision (GN-Vision) system, which aims at simultaneously obtaining information about the spatial origin of \g-ray and neutron sources. We first investigated the performance of two large 50$\times$50~mm$^{2}$ monolithic CLYC crystals, 8 and 13~mm thick respectively, coupled to a pixelated SiPM in terms of energy resolution and neutron-gamma discrimination. The response of two different 95\% $^{6}$Li-enriched CLYC detectors coupled to an array of 8$\times$8 SiPMs was studied in comparison to the results of a conventional photo-multiplier tube. An energy resolution of about 6\% with PMT and 8\% with SiPMs for the $^{137}$Cs peak and a figure of merit of 3-4 for the neutron-gamma discrimination have been obtained. The spatial response of the CLYC-SiPM detector to $\gamma$-rays and neutrons has also been characterized using charge modulation-based multiplexing techniques based on a diode-coupled charge division circuit. Average resolutions close to 5~mm FWHM with good linearity are obtained in the transverse crystal plane. Last, this work presents the first proof-of-concept experiments of the neutron imaging capability using a neutron pinhole collimator attached to the developed position sensitive CLYC detector.
\end{abstract}

\begin{keyword}
%Dual neutron-gamma imaging, Neutron imaging, Hadron therapy, BNCT, Nuclear inspections, Special nuclear material, Position-sensitive detectors, Pulse shape discrimination, CLYC, SiPM, Monolithic crystals, multiplexing methods.
Dual neutron-gamma imaging; Neutron dosimetry; Nuclear inspections; Position-sensitive detectors; Monolithic CLYC; SiPM.
\end{keyword}

\end{frontmatter}

%\linenumbers

\section{Introduction}

%\begin{itemize}
 %   \item Main motivation medical application
  %  \item Additional motivation for nuclear inspections
  %  \item State-of-the art and limitations
  %  \item Solutions with GN-Vision
   % \item Introduce scheme of the article
%\end{itemize}
The deployment of innovative detection technologies capable of identifying special nuclear material (SNM)~\cite{Polack:11,Poitrasson:15,Petrovic:21}, controlling spent fuel in reactors~\cite{Parker:15} or inspecting nuclear accidents unmanned~\cite{Sato:19,Vetter:18} is of key strategic importance. In light of this, simultaneous real-time imaging of $\gamma$-rays and neutrons has become increasingly relevant, since it allows the location of a broader range of radioactive sources. Most of the deployed devices combining neutron and gamma imaging to date were based on large arrays of liquid scintillation detectors \cite{Pozzi:12,Poitrasson:14}, which are sensitive only to fast neutrons and have limited portability and applicability due to the bulk of the system. Due to the advantages of such multi-mode imaging devices for such applications, several groups have been working in recent years on the development of more compact dual imaging devices~\cite{Soundara:12,Whitney:15,Hamrashdi:20,Steinberger:20,Boo:21,Guo:21,Lopez:22}.

The application of dual neutron-gamma imaging devices also holds significant promise in the field of proton and ion beam therapy. These treatment techniques address two major challenges: real-time dose monitoring for neutrons and gamma rays~\cite{Schneider:15}, and ion-beam range verification~\cite{Durante:19}. These issues currently limit the advantages of proton therapy compared to photon therapy. Dual neutron-gamma imaging prototypes offer a promising solution to these challenges; however, the size of many existing devices poses a constraint for their use in clinical treatment rooms~\cite{Clarke:16}. Recent conceptual studies have explored the development of compact dual imaging devices specifically designed for range verification~\cite{Meric:23,Schellhammer:23}. Neutron and gamma imaging can also provide valuable information for accurate dosimetry in Boron Neutron Capture Therapy (BNCT)~\cite{Hou:22}.

In this context, we have designed a new dual neutron- and \g-ray-imaging tool~\cite{Patent}, hereafter referred as GN-Vision, that aims to address the most relevant challenges for the aforementioned applications. 
 The system consists of a compact and handheld-portable device capable of measuring and simultaneously imaging \g-rays and slow -- thermal to 100~eV -- neutrons, both of them with a high efficiency~\cite{Lerendegui:22_ANPC,Lerendegui:24}. The proposed device consists of an upgrade of the Compton imager i-TED (Imaging Total Energy Detector) ~\cite{Domingo16}, a device that has been fully developed at Instituto de F\'isica Corpuscular (IFIC)~\cite{Olleros18,Babiano19,Babiano20,Balibrea:21} and used over the last years for experiments of astrophysical relevance at the CERN Neutron Time of Flight (n\_TOF) Facility~\cite{Babiano:21,Lerendegui:23_NPA,Domingo:23}. Moreover, the applicability of i-TED to range verification in ion beam therapy ~\cite{Lerendegui:22,Balibrea:22} and imaging-based dosimetry in BNCT~\cite{Lerendegui:24_AppRadIsot,Torres:24} has been explored with promising results, as well as its application to nuclear waste characterization~\cite{Babiano:24}.
 
GN-Vision was conceptually designed on the basis of Monte Carlo simulations which demonstrated its simultaneous \g-ray and neutron detection and imaging capabilities~\cite{Lerendegui:22_ANPC,Lerendegui:24}. For the imaging of \g-rays, this device operates as a Compton camera consisting of two position sensitive detection planes. In order to achieve the imaging of slow neutrons~\cite{Lerendegui:24}, the active material of the first position sensitive detection layer of GN-Vision is chosen to be a CLYC detector, with the capability of discriminating \g-rays and neutrons. The second plane, acting as the absorber of a Compton camera, comprises four LaCl$_3$ crystals as in the predecessor i-TED device. A passive pinhole collimator made of a material with high absorption power for thermal and epithermal neutrons  attached to the first detection plane allows one to perform neutron imaging with the same working principle as pinhole cameras~\cite{Anger:58}. The collimator is made from a lightweight material, such that interference with \g-rays is very low.

The central component of GN-Vision is its first position sensitive detection layer featuring particle discrimination capability. A monolithic Cs$_{2}$LiYCl$_{6}$:Ce scintillation crystal enriched with $^{6}$Li at 95\% (CLYC-6), able of unambiguously separating \g-rays, fast and thermal neutrons by Pulse Shape Discrimination (PSD)~\cite{Giaz:16}, was chosen for this purpose~\cite{Lerendegui:24}. Slow neutrons reaching the CLYC-6 interact via the $^{6}$Li(n,$\alpha$)$^{3}$H reaction. The outgoing tritium and alpha particles deposit about 4.78~MeV in the crystal, which result in a light production of 3.2~MeVee~\cite{Giaz:16}. Fast neutrons can also be detected via the $^{35}$Cl(n,p) and $^{35}$Cl(n,$\alpha$) reactions~\cite{Giaz:16}. In recent years, CLYC crystals have received an increasing attention due to the combined detection properties and good resolution for its use mostly in neutron-gamma monitoring devices~\cite{Budden201597,Plaza:23,Qi:23} and more scarcely in imaging devices~\cite{Soundara:12,Whitney:15}.

In this work we present the first experimental milestones in the deployment of GN-Vision, focused on the development and characterization of the position-sensitive CLYC detector and the neutron imaging capability. In Sec.~\ref{sec:CLYC}we present the performance in terms of energy resolution and particle discrimination of several CLYC crystals coupled to a photo-multiplier tube (PMT) and a pixelated silicon photo-multiplier (SiPM). Sec.~\ref{sec:Position_reconstruction} deals with the characterization of the spatial response of the CLYC-SiPM detector and the first test of position-sensitive neutron-gamma discrimination. Following the latter achievement, in Sec.~\ref{sec:NeutronImaging} we present the first proof-of-concept experiment of the neutron imaging capability of GN-Vision. Last, a summary of our results and the outlook for the development of the complete GN-Vision prototype are provided in Sec.~\ref{sec:Summary}.

\section{Performance of the CLYC detectors}\label{sec:CLYC}

\subsection{CLYC crystals and readout electronics}\label{sec:CLYC_Exp}

The central component of the dual imaging device, GN-Vision, is the first position sensitive detection layer with \g-ray and neutron discrimination capability~\cite{Lerendegui:24}, realized with a CLYC-6 crystal, hereafter called simply CLYC.

In the first part of the experimental characterization of the first position sensitive detection plane, the performance of two different CLYC crystals, hereafter referred to as CLYC-A (RMD, S/N: CLYC-1092-4) and CLYC-B (Capesym, S/N:~010507), with dimensions of 50$\times$50$\times$8 mm and 50$\times$50$\times$13 mm, respectively, was compared in terms of energy resolution and neutron-gamma discrimination capability. Both crystals were sealed by the provider in an aluminum housing and a 2~mm thick quartz window. In order to compare the two crystals, the same PMT, Hamamatsu R6236 with a rectangular window of 60$\times$60 mm was coupled to the optical window of the CLYC crystals with the help of silicone optical grease (Saint Gobain BC-631). This PMT exhibits a spectral response spanning from 300 to 650 nm, rendering it suitable for detecting the 370 nm scintillation photons emitted by the CLYC crystal. %The PMT was operated at a voltage of 1~kV, beyond which no gain in resolution was found.

With the aim of deploying an imaging system, a position sensitive CLYC detector is required~\cite{Lerendegui:24}. For this reason, further developments with a SiPM were pursued. Following the initial assessment of the detector's performance with PMT readout, the best suited CLYC was coupled to the OnSemi SiPM ARRAYJ-60035-64P, the same model used in the predecessor i-TED detector~\cite{Babiano19,Balibrea:21}. This photosensor features 8$\times$8 pixels over a surface of 50$\times$50 mm$^{2}$. 
 %Moreover, it becomes convenient to use a low-volume photosensor for the CLYC-detector in order to enable the assembly of the full Compton module.
The CLYC-SiPM detector was read out with two different types of electronics. In a first comparison with the PMT readout, the SiPM output was connected to an ARRAYX-BOB6-64S sum board. This breakout board allows easy access to the summed output, achieved by connecting the anodes of pixels 1 to 64 together of all standard pixel signals. The output of this board mimics the single output of a PMT, which is sufficient to evaluate the performance of the CLYC crystal in terms of energy resolution and pulse-shape discrimination. To enable the position sensitivity, discussed in Sec.~\ref{sec:Position_reconstruction}, the final readout of the CLYC-SiPM module was carried out with the AIT 4-Channel Active Base (AB4T-ARRAY64P)~\cite{AIT}, designed for the readout of the Onsemi SiPM Array J-60035-64P-PCB. This electronic device is based on a diode-coupled charge division readout, which features superior performance compared to traditional resistive readouts. In particular, it provides improved spatial uniformity, faster rise time, and reduced image noise. Employing charge modulation-based multiplexing (i.e. Anger Logic)~\cite{Anger:58}, it provides four encoded position signals for event centroid calculation out of the 64 signals of the individual pixels (see Sec.~\ref{sec:Pos_resol}). This device provides a total of five output channels, comprising the aforementioned four weighted position signals and a summed signal, which provides the information about the total charge collected, allowing this system to perform spectroscopy and pulse shape discrimination as well. The results of the CLYC-SiPM device presented hereafter correspond only to the 4-Channel Active Base readout.

The output signals of the PMT and the AB4T-ARRAY64P were fed into a CAEN DT5730S digitizer, which reads out signals with 14-bit resolution Flash ADCs at a 500 MS/s sampling rate. This data acquisition (DAQ) system was operated in online digital pulse processing (DPP) mode, in which two integration windows (short and long gate) are defined to enable pulse-shape discrimination (PSD), as it is explained in Sec.~\ref{sec:CLYC_NeutGamma}. The long and short integrals for each signal are saved in time-stamped list mode. 

% \begin{itemize}
 %   \item Detector properties, providers, etc..
%    \item photosensors: PMT, SiPM, sum PCB
 %   \item CAEN DAQ
  %  \item AIT
  %  \item Figures: show (maybe) a sketch of the setup
%\end{itemize}

\subsection{Energy resolution}\label{sec:CLYC_Energy}
%\begin{itemize}
 %   \item Compare performance of two crystals with PMT
  %  \item Performance of SiPM vs PMT
   % \item Show one spectrum and (maybe) one optimization curve
%\end{itemize}

\begin{figure}[!htbp]
\centering
\includegraphics[width=0.9\linewidth]{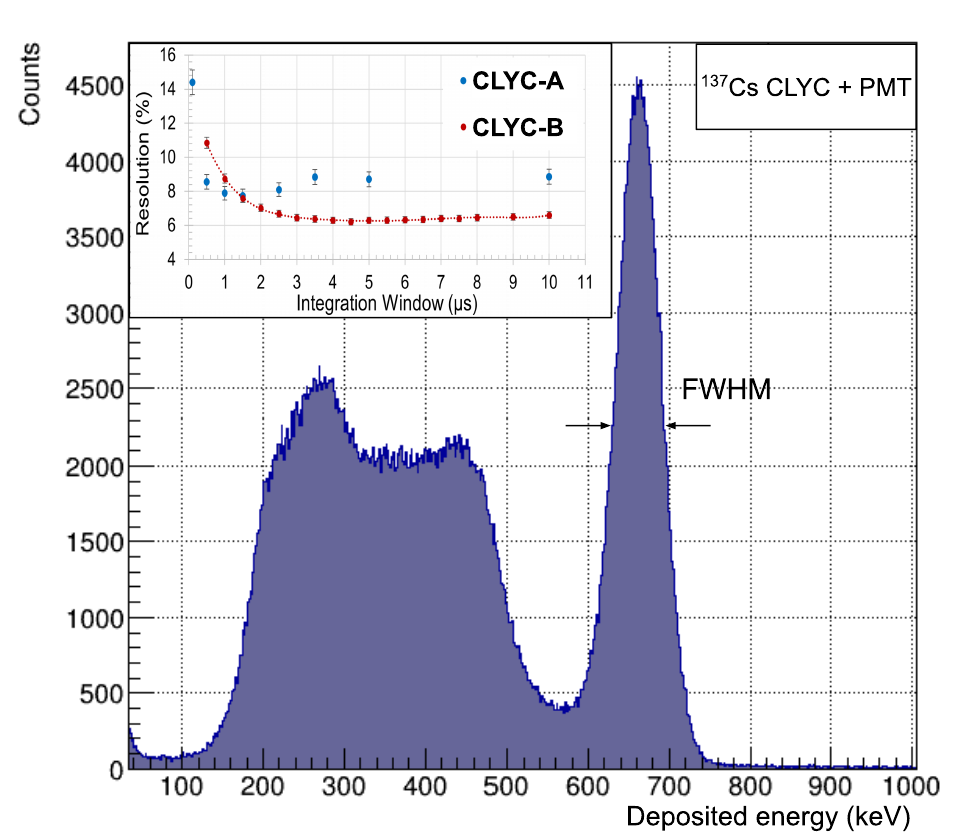}
\caption{Deposited energy spectra of the CLYC-B scintillator coupled to the PMT. The energy resolution, computed from the FHWM of the 662 keV photo-peak, as a function of the integration window for the two CLYC scintillators is shown in the insert.}
\label{fig:Spectrum_resolutiion}
\end{figure}

The performance of the two CLYC detector models was first studied in terms of energy resolution. For this purpose, the 662~keV photo-peak registered during the measurement of a $^{137}$Cs source was analyzed. As an example, Fig.~\ref{fig:Spectrum_resolutiion} shows the energy-calibrated spectrum registered with the CLYC-B detector coupled to the Hamamatsu PMT R6236 PMT. 

The dependence of the energy resolution with the operation HV of the PMT and the integration window (long gate) of the CAEN DAQ was evaluated. The resolution was found to be independent of the PMT HV above 0.9 kV and consequently, a voltage of 1 kV was set for the rest of the measurements. As for the integration window, the top left inset of Fig.~\ref{fig:Spectrum_resolutiion} shows that for the CLYC-A crystal an optimum resolution of 7.5\% is found for integration gates of 1.5~$\mu$s, while in the case of the CLYC-B crystal it keeps improving for integration windows up to 3~$\mu$s. The latter detector is found to have a significantly better resolution of 6.2\%. When compared with the best values reported in the literature with PMT readout, which are as good as 4\% to 5.5\% \cite{West20,Lee12,Qin18,Wang22}, the resolution in this work (6.2\%) seems significantly worse at first glance. However, all the previous works use of spectroscopic amplifiers for an optimized energy resolution. Indeed, the providers of the two CLYC detectors report energy resolutions 5.0\% (CLYC-A) and 4.2\% (CLYC-B) using PMT readout, preamplifier and spectroscopic amplifier. In contrast, the present investigation is aimed towards particle discrimination with PSD which requires the digitization of the raw signal. The obtained results for CLYC-A and CLYC-B, should not be regarded as representative of the different crystal manufacturers, as these are just two specific measurements that may be affected by different experimental aspects such as the optical coupling, the choice of PMT, etc.

The CLYC-B crystal, exhibiting the best resolution with the PMT, was selected for the characterization with the SiPM readout. A similar experimental procedure was conducted for the summed readout of the SiPM. Upon varying the integration window from 1.5 to 6 microseconds, the best resolution was found with an integration window at 3~$\mu$s, yielding a value of 8.9\%. The worsening of the resolution with the SiPM array with respect to the PMT readout could be ascribed to the worse signal-to-noise ratio obtained with the former. Compared to similar studies in the literature, our result is better or comparable to values reported in previous works such as~\cite{Wang19,Qi:23}. Although better results have been reported with SiPMs in the literature~\cite{West20,Huang21}, they involve external amplification techniques that eliminate potential position sensitivity, rendering them unsuitable for the overall imaging objective.

\subsection{Neutron-gamma discrimination}\label{sec:CLYC_NeutGamma}

The possibility of using pulse shape discrimination (PSD) to distinguish the incoming $\gamma$-rays from neutrons is the key feature of the CLYC for its use in the dual gamma-neutron device. The discrimination is based on the differences in the scintillation decay response. While $\gamma$-ray events manifest as sharp, quick rises in the pulse, indicative of rapid light production, neutron events exhibit broader pulses with slower decay~\cite{Giaz:16}. The disparity in pulse shapes can be quantified, following the procedure in the literature~\cite{Olympia:13,Giaz:16}, by integrating each pulse within two non-overlapping windows W1 and W2, which integrate, respectively, the prompt and delayed parts of the signal. Assuming that the signal integrals in these windows are $Q_{\text{1}}$ and $Q_{\text{2}}$ respectively, the PSD ratio, used to perform the discrimination, is defined as:

\begin{equation}\label{eq:PSD}
\text{PSD} = \frac{Q_2}{Q_1 + Q_2}
\end{equation}

\begin{figure}[!b]
\centering
\includegraphics[width=0.9\linewidth]{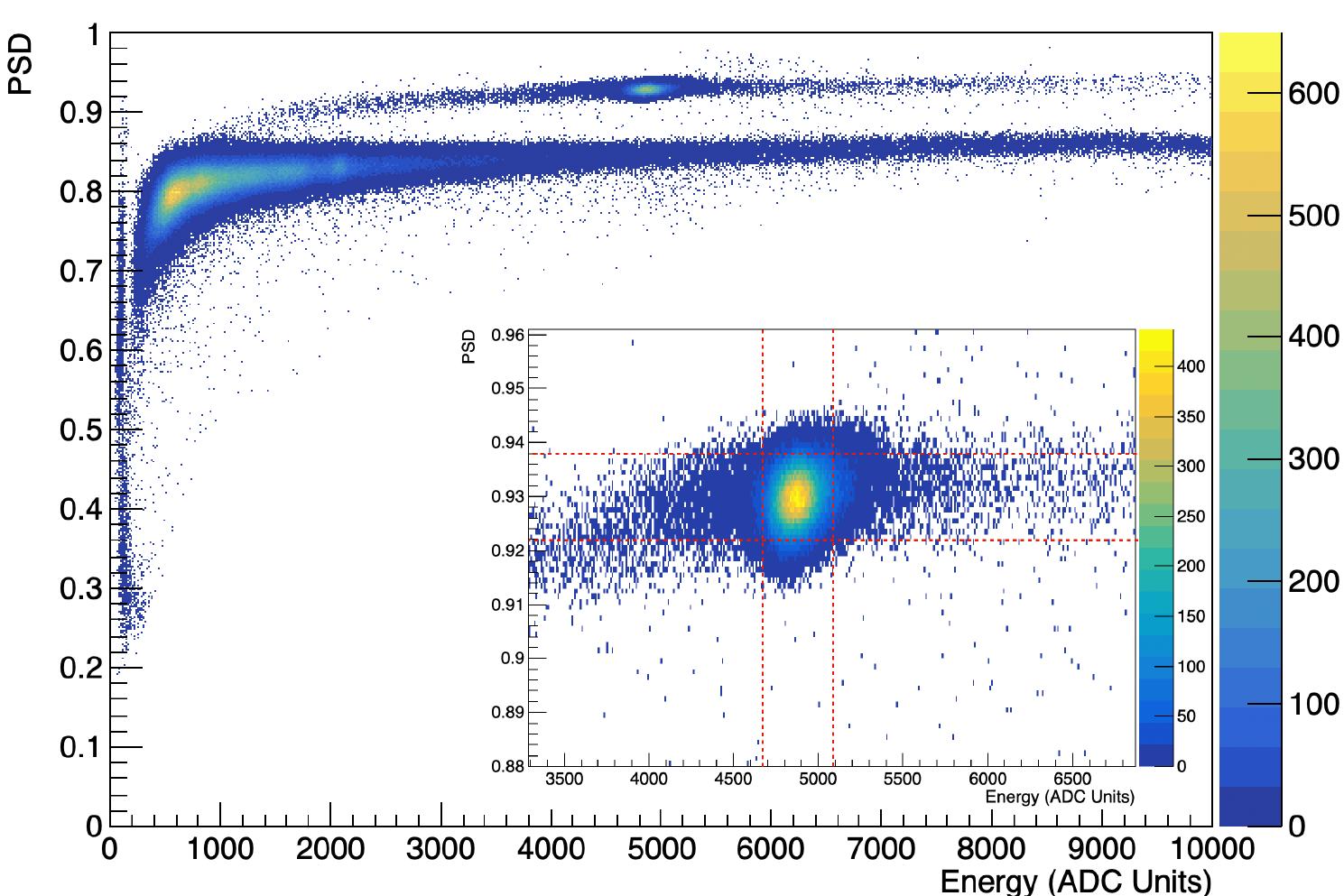}
\caption{PSD as a function of the energy (ADC channel) for the CLYC-B coupled to a SiPM array obtained in an irradiation with a partially moderated $^{252}$Cf source.}
\label{fig:PSD_ADC_SiPM}
\end{figure}

\begin{figure}[!htbp]
\centering
\includegraphics[width=0.9\linewidth]{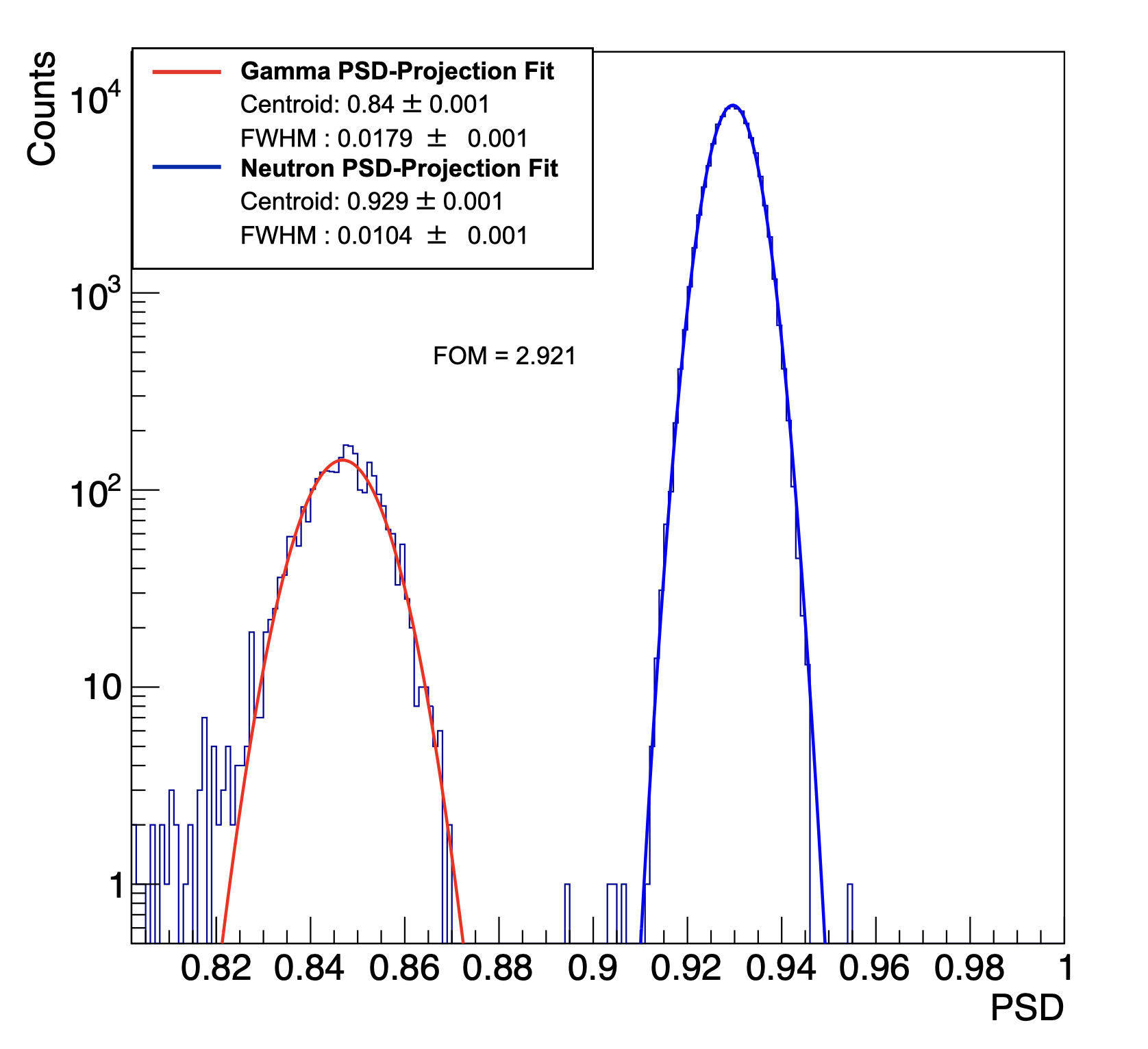}
\caption{Projection of Fig.~\ref{fig:PSD_ADC_SiPM} onto the PSD axis of the neutron and gamma-ray distributions with a cut of 2.5-4 MeV around the neutron bump. A FOM of 2.91 has been obtained from the analysis for the CLYC-B coupled to a SiPM array.}
\label{fig:FOM_Final_SiPM}
\end{figure}

In a similar fashion to the characterization of the energy resolution, we first evaluated the PSD performance of both the CLYC-A and CLYC-B crystal coupled to the Hamamatsu PMT R6236 and then, only for the CLYC-
B crystal, with SiPM readout. For this study, a $^{252}$Cf source, neutron emitter via spontaneous fission, was embedded in a Polyethylene (PE) matrix of 5~cm thickness to partially moderate the neutrons~\cite{Lerendegui:24}. The detectors were attached to the PE surface to maximize the efficiency. Two integration windows, so-called long gate (LG=W2+W1) and short gate (SG=W1), were defined in the CAEN DAQ settings to integrate each of the registered signals and to generate 2D matrices of PSD vs amplitude (i.e. deposited energy). Fig.~\ref{fig:PSD_ADC_SiPM} shows the resulting matrix for the result obtained with the CLYC-SiPM detector. The inset in this figure shows a zoom of the neutron branch, which shows a clear bump, corresponding to the detection of thermal and epithermal neutrons via the $^{6}$Li(n,$\alpha$)$^{3}$H reaction, located at 3.2 MeVee~\cite{Giaz:16}.  

To evaluate the quality of the neutron-gamma discrimination, we calculate a Figure of Merit (FOM) commonly found in the literature~\cite{Giaz:16}. Selecting in the PSD-energy histogram events within the vertical dashed lines shown in the inset of Fig.~\ref{fig:PSD_ADC_SiPM} ($\pm 2\sigma$ around the maximum) and projecting them on the PSD axis, we obtain a spectrum with two well-separated peaks: one related to neutron hits (higher PSD) and the other to $\gamma$-ray hits (lower PSD), which are shown in Fig.~\ref{fig:FOM_Final_SiPM}. From the projection, it is possible to extract the FOM, defined as:

\begin{equation}\label{eq:FOM}
\text{FOM} = \frac{|\mu_n - \mu_\gamma|}{\text{FWHM}_\gamma + \text{FWHM}_n}
\end{equation}

 where ${\mu}_n$ and ${\mu}_\gamma$ are, respectively, the centroids of the neutron and $\gamma$-ray peak, while $\text{FWHM}_n$ and $\text{FWHM}_\gamma$ are the corresponding peak widths; see Fig.~\ref{fig:FOM_Final_SiPM}. 

\begin{figure}[!t]
\centering
\includegraphics[width=0.9\linewidth]{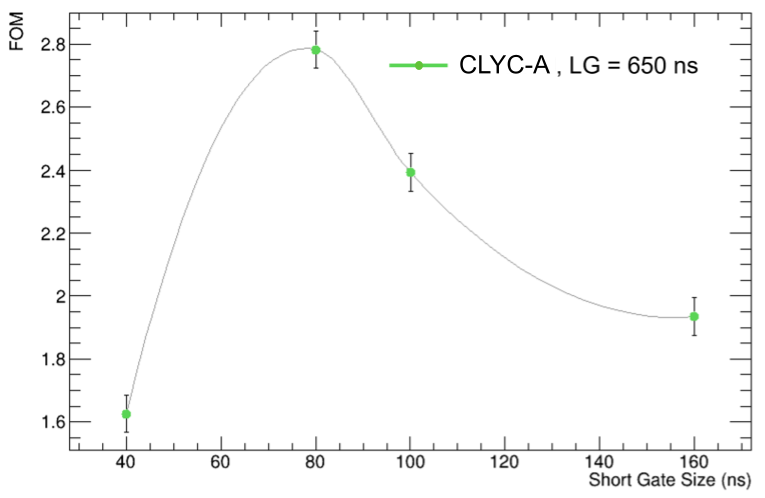}
\includegraphics[width=0.94\linewidth]{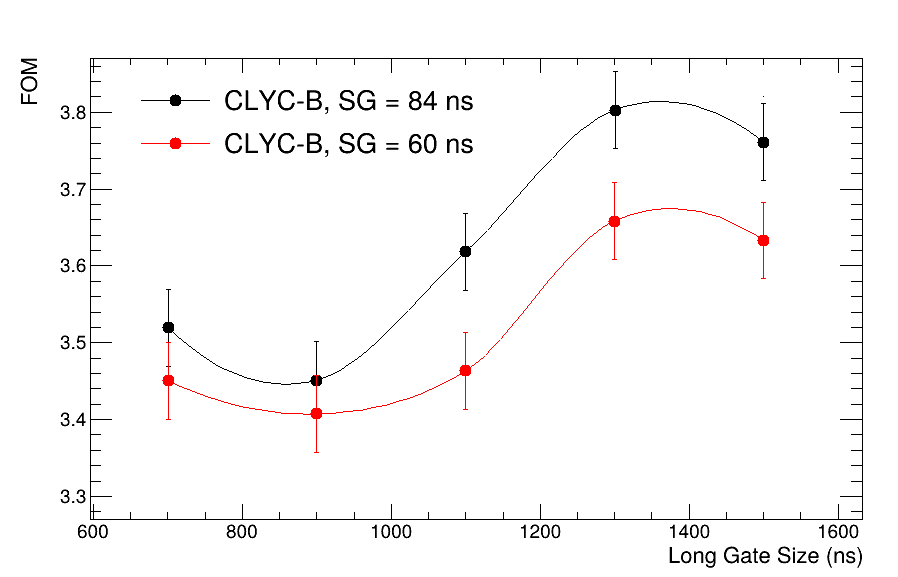}
\caption{Figure of Merit of the PSD depicting the optimum integration windows for the two CLYC models. Top: SG varied with CLYC-A, revealing a clear peak at 80~ns. Bottom: LG altered with CLYC-B, showing a clear peak around 1400 ns. A consistent improvement was observed for SG values with 84 ns compared with 60 ns. The curves are smooth lines connecting the data points.}
\label{fig:FOM_Crystals_PMT}
\end{figure}

In a similar way to the energy resolution, the long and short integration gates (LG and SG) of the CAEN DAQ were optimized in terms of the FOM of the PSD. The results for the two CLYCs coupled to the PMT, shown in Fig.~\ref{fig:FOM_Crystals_PMT}, indicate that an optimum FOM was found for a SG of around 80~ns for both crystals. As for the LG, the CLYC-A showed the best FOM with LG = 650 ns, whereas a distinct rise was observed for CLYC-B at LG = 1400 ns. The best FOM values of $2.8\pm0.05$ and $3.8\pm0.05$ were obtained for the CLYC-A and CLYC-B, respectively. The systematic uncertainties in Fig.~\ref{fig:FOM_Crystals_PMT} represent the maximum observed fluctuation of the FOM value resulting from choice of cut around the thermal neutron bump in the analysis.
The PSD FOM was also studied for the CLYC + SiPM assembly -- only for the CLYC-B, which outperformed CLYC-A both in energy resolution and PSD performance-- yielding a best FOM of 2.92, corresponding to the results of Fig.~\ref{fig:FOM_Final_SiPM}.

Our best FOM of 3.8, obtained with the PMT and the CLYC-B crystal, lies in the upper range of the literature values (2.9 to 4.55 \cite{Higgins10,Lee12,West20,Qin18,Soundara:17,Wang22,Plaza:23}). The SiPM yielded a FOM of 2.9, which is comparable to or better than previous results reported for a similar setup, ranging from 1.8 to 2.8 \cite{West20,Dinar19,Wang19}. Higher FOM values, up to 4, have been reported with SiPMs that employ signal amplification techniques \cite{Huang21}. 

\section{Position resolution}\label{sec:Position_reconstruction}

As has been shown for the predecessor i-TED detector~\cite{Balibrea:21}, the Compton $\gamma$-ray imaging performance depends, in a different measure, on the accuracy of the determination of the $\gamma$-ray hit position along the transverse plane and -- in the case of thick crystals -- the depth of interaction. In the case of neutron imaging, GN-Vision acts as a simple pinhole camera~\cite{Lerendegui:24}, where the image resolution is directly correlated to the intrinsic spatial resolution of the CLYC crystal coupled to the SiPM. Therefore, it is important to characterize the spatial resolution of the CLYC with SiPM readout, ensure the best possible performance and eventually correct for non-linearity distortions particularly in the peripheral region of the crystal volume, so-called pin-cushion effect~\cite{Balibrea:21}. For the case of the CLYC detector, the position sensitivity has to be combined with the capability to discriminate the incoming radiation, thus enabling position-sensitive neutron-gamma discrimination.

\subsection{Experimental setup}\label{sec:Pos_Exp}
\begin{figure}[!htbp]
\centering
\includegraphics[width=0.92\linewidth]{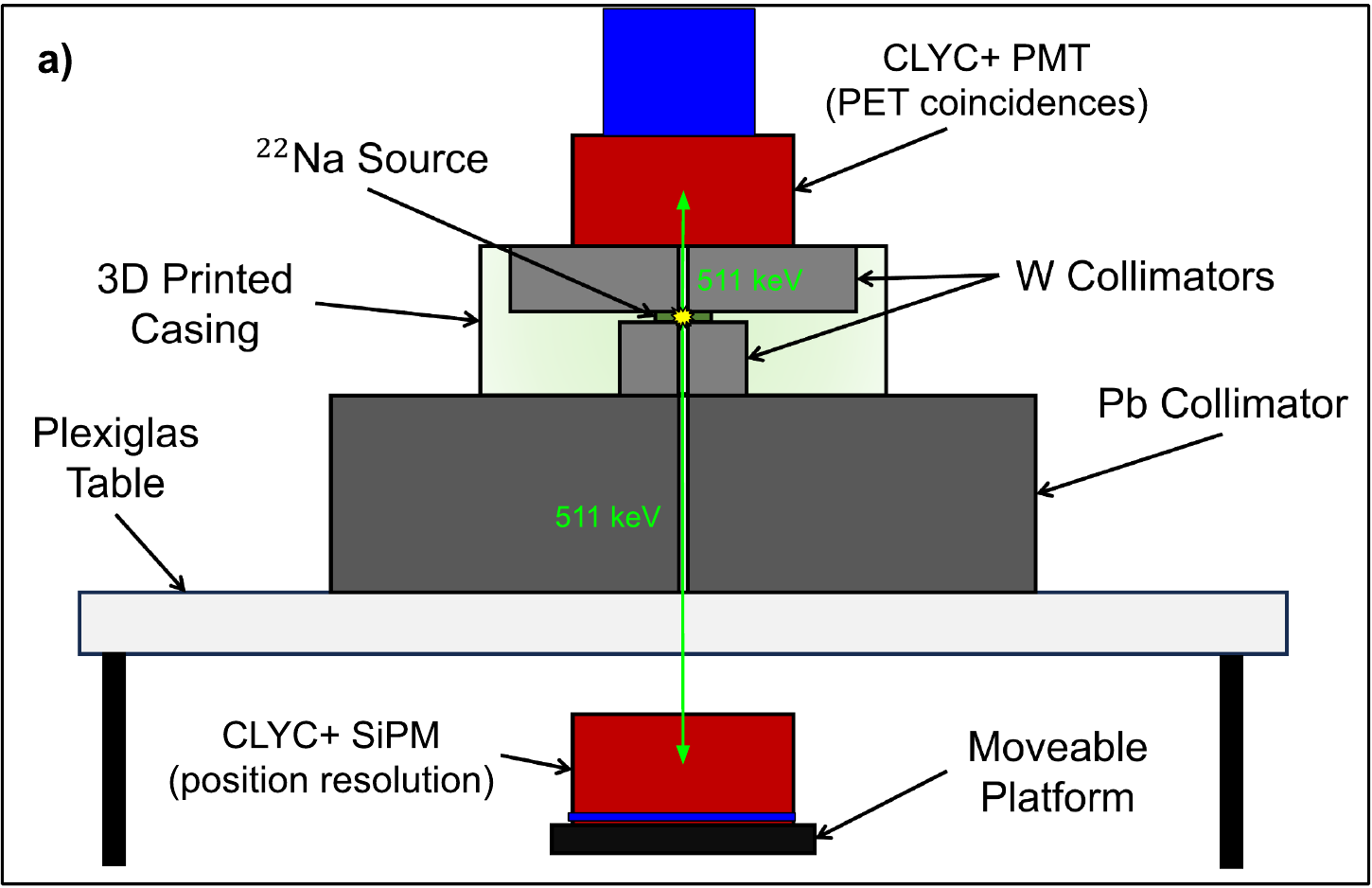}
\includegraphics[width=0.93\linewidth]{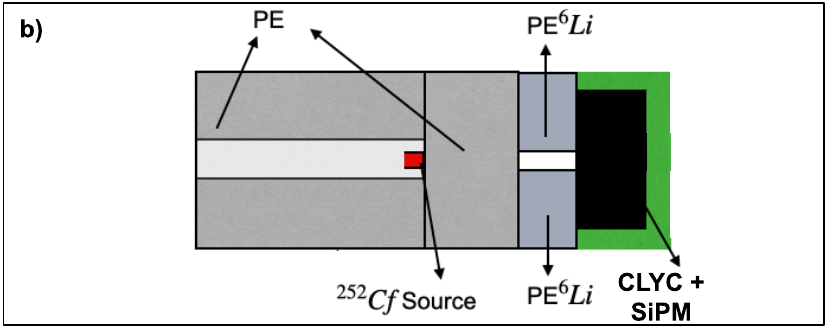}
\caption{a) Sketch of the $\gamma$-ray scanning table using a collimated $^{22}$Na source indicating the main components. b) Experimental setup based on a partially moderated $^{252}$Cf source used to validate the reconstruction of neutron patterns. More details are given in the text }
\label{fig:Setup_Pos}
\end{figure}

The performance of the CLYC coupled to the SiPM array and read out with the 4-channel active base was evaluated in terms of resolution, linearity, and compression with the help of a $\gamma$-ray scanning table, sketched in the top panel of  Fig.~\ref{fig:Setup_Pos}. In a second step, the position-sensitive neutron-gamma discrimination capability was proved with help of a moderated neutron source and a slit in a neutron-absorbing material, shown in the bottom panel of the same figure. The details are explained in the following. 

The systematic spatial characterization in the $x$-$y$ transverse crystal plane was carried out using a $\gamma$-ray scanning table based on a collimated beam of $\gamma$-rays in combination with an $x,y$-positioning table. This experimental setup is schematically shown in Fig.~\ref{fig:Setup_Pos} and is similar to the one used in Ref.~\cite{Balibrea:21}.  The position-sensitive detector under characterization is sitting in a small movable platform at the bottom of the experimental setup, whereas the collimating system, the $^{22}$Na-source, and the subsidiary detector are placed and fixed on top of a Plexiglas table. The $x,y$-gantry which drives the movable platform is the T-G-LSM200A200A model from Zaber Technologies, which has an accuracy of 32~$\mu$m and repeatability of 3~$\mu$m~\cite{Balibrea:21}. The $^{22}$Na-source is sandwiched in a collimating structure under the subsidiary detector to achieve a pencil beam in the direction perpendicular to both crystal surfaces. The upper part of this collimator is made of a tungsten parallelepiped with a central hole of 1~mm and 30~mm thickness. Under the radioactive sample a circular tungsten collimator is placed with a central hole of 1~mm and 24~mm effective thickness. An additional lead collimator of 5.5~cm thickness is placed beneath the collimating structure. The aperture of this lead collimator is of 3~mm diameter. The CLYC-SiPM detector under characterization (bottom) is operated in time-coincidence mode -- 15~ns coincidence window -- with a second CLYC-PMT detector placed in the top part of the setup. In total, the distance between the front face of both CLYC crystals in the experimental setup is 26~cm. For the analysis we use only full-energy deposition of the 511~keV annihilation $\gamma$-rays emitted from a point-like $^{22}$Na-source (1~MBq activity), thus removing possible background events not associated with the collimated 511~keV $\gamma$-ray beam.

As for the setup used to validate the position-sensitive neutron-gamma discrimination (bottom panel of Fig.~\ref{fig:Setup_Pos}), the same  \(^{252}\text{Cf}\) source used for the PSD characterization in Sec.~\ref{sec:CLYC_NeutGamma} was surrounded by a 5-cm-thick matrix of PE acting as a moderator of the fission neutron spectrum. In the SF decay of \(^{252}\text{Cf}\) not only neutrons but also $\gamma$-rays are emitted. To evaluate the position sensitivity to neutrons, a simple pattern consisting of a 5~mm opening slit made with 2~cm thick layers of $^6$LiPE of 95\%enriched $^{6}$LiPE was placed just after the exit of the neutron moderator. The very same neutron absorbing material has been also used for the neutron pinhole collimator of GN-Vision~\cite{Lerendegui:24}. 

The SiPM readout electronics and DAQ, introduced in Sec.~\ref{sec:CLYC_Exp}, have been used for the position-sensitive measurements of this section. The details on the position reconstruction methodology are given in Sec.~\ref{sec:Pos_resol}

\subsection{Spatial linearity and resolution}\label{sec:Pos_resol}

\begin{figure*}[!htbp]
\centering
\includegraphics[width=0.90\linewidth]{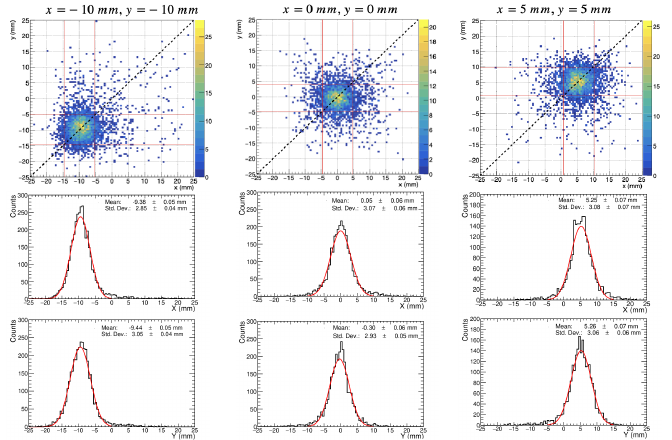}
\caption{Reconstructed positions for different irradiation points on the CLYC crystal surface and projections onto the$\times$and Y axes to fit the total Gaussian broadening. The dotted black line indicates the diagonal for visual
reference, and the red lines represent the cuts applied to obtain the projections.}
\label{fig:Positions}
\end{figure*}

Using the $\gamma$-ray scanning system described in Sec.~\ref{sec:Pos_Exp}, measurements of 12~h were taken in 30 irradiation positions covering the Y and X axes, as well as the two crystal diagonals~\cite{GastonThesis}. After both the mechanical collimation and the coincidence filtering method with the two CLYCs, we selected only events of the 511 keV $\gamma$-ray pencil-beam for the position reconstruction. For each of the selected $\gamma$-ray hits in the CLYC-SiPM detector we reconstructed the X and Y coordinates of each from the integral of the four position-encoded signals ($X_+$, $X_-$, $Y_+$, $Y_-$) generated by the AB4T-ARRAY64P readout electronics, as follows:
\begin{equation}\label{eq:codex}
\begin{aligned}
    X=\frac{(X_+ - X_-)}{(X_+ + X_-)},~~~
    Y=\frac{(Y_+ - Y_-)}{(Y_+ + Y_-)}
\end{aligned}
\end{equation}

Fig.~\ref{fig:Positions} shows the 2D distribution of reconstructed positions for three different irradiation positions along the diagonal of the CLYC crystal. In this figure, one can appreciate clear patterns of reconstructed positions of interaction around the maxima. The X and Y coordinates of the reconstructed position and the spatial resolution in both axes are obtained from the Gaussian fit of the projection of the 2D distributions onto the X and Y axes, as shown in the bottom panel of Fig.~\ref{fig:Positions}.

The experimental distribution of reconstructed positions in Fig.~\ref{fig:Positions} is noted in the following as \(f(\sigma_{\text{MEASURED}},\mu)\), where $\sigma_{\text{MEASURED}}$  and $\mu$ represent, respectively, the measured standard deviation and the centroid. This distribution is a convolution of the intrinsic response of the CLYC-SiPM assembly with the AIT 4-channel active base readout and the beam divergence due to the non-ideal collimation of the $\gamma$-ray scanning table~\cite{Balibrea:21}. To isolate the spread of the reconstructed positions due to the non-punctual pencil-beam in our experimental setup, given by the distribution \(g(\sigma_{\text{SETUP}},\mu)\), from the inherent resolution of the CLYC-SiPM detector under characterization, characterized by the distribution \(\mathcal{N}(\sigma_{\text{CLYC}},\mu)\), we determined the contribution of the collimation system \(g(\sigma_{\text{SETUP}},\mu)\) from a Monte Carlo simulation carried out with the \textsc{Geant4} toolkit~\cite{Geant4_2}, where the precise geometry of the setup was modelled. The final aim of this MC study is to determine the interplay between the intrinsic detector spatial resolution and the overall (measured) width. The latter can be graphically represented in order to deconvolve the intrinsic resolution of interest, as has been performed in previous works~\cite{Babiano19,Balibrea:21}. The results of these MC simulations are used to determine \(\mathcal{N}(\sigma_{\text{CLYC}},\mu)\) from the relationship:

\begin{equation}\label{eq:convolution}
f(\sigma_{\text{MEASURED}}, \mu) = g(\sigma_{\text{SETUP}}, \mu) \circledast \mathcal{N}(\sigma_{\text{CLYC}}, \mu).
\end{equation}

\begin{figure}[!h]
\centering
\includegraphics[width=0.8\linewidth]{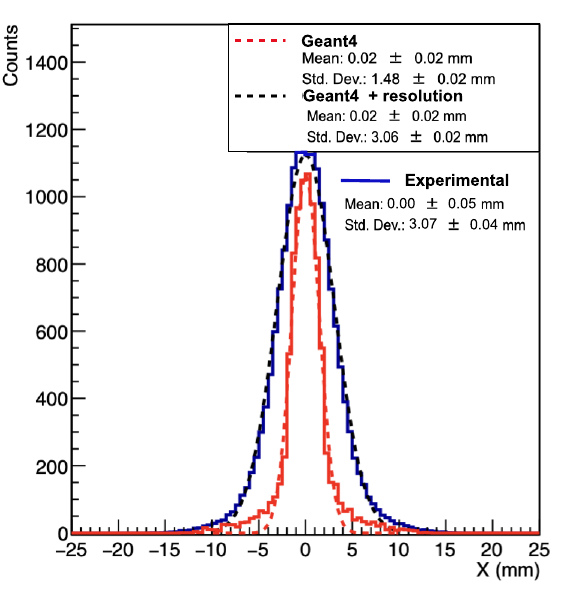}
\caption{Projection onto the X axis of the spatial distribution of $\gamma$-rays impinging on the CLYC crystal. The ideal distribution extracted from the simulations of the setup (red), which accounts only for the non-ideal collimation, has been broadened with a Gaussian resolution (dashed black) so that the resulting distribution matches the width of the experimental distribution (blue).}
\label{fig:Geant4_resolution}
\end{figure}

The \textsc{Geant4} simulation paid special attention to the sensitive distances and the hole diameter of the different collimators shown in Fig.~\ref{fig:Setup_Pos}. The realistic decay of the $^{22}$Na source was simulated making use of G4RadioactiveDecay, followed by the positron-electron annihilation that gives rise to the two 511 keV $\gamma$-rays. Only one position, with the source in the center of the CLYC-SiPM detector, was simulated and the position of the electromagnetic interactions of the collimated $\gamma$-ray beams was registered for decays where both CLYC crystals were fired by a 511~keV $\gamma$-ray to resemble the experimental data set. The registered XY distributions were extracted and projected onto the X and Y axes. The spatial distribution from the simulation, accounting only for the divergence of the beam, is shown in red in Fig.~\ref{fig:Geant4_resolution}. Only the X coordinate is shown because of the symmetry of the collimator. In order to match the experimental distribution, represented by the blue curve, the $\gamma$-ray interaction points from the MC simulations were convoluted with a Gaussian resolution, resulting in the dashed black curve. For the central position, shown in Fig.~\ref{fig:Geant4_resolution}, the standard deviation of the convoluted projections are \(\sigma_x = 3.06(2)\) mm and \(\sigma_y = 3.05(2)\) mm, very close to the measured projections are \(\sigma_x = 3.07(6)\) mm and \(\sigma_y = 2.93(5)\) mm, respectively. The resulting unfolded intrinsic resolution for the central position is 2.25~mm. This methodology was followed to extract the intrinsic resolution in the X and Y coordinate individually for all the irradiation positions.

\begin{figure}[!t]
\centering
\includegraphics[width=0.85\linewidth]{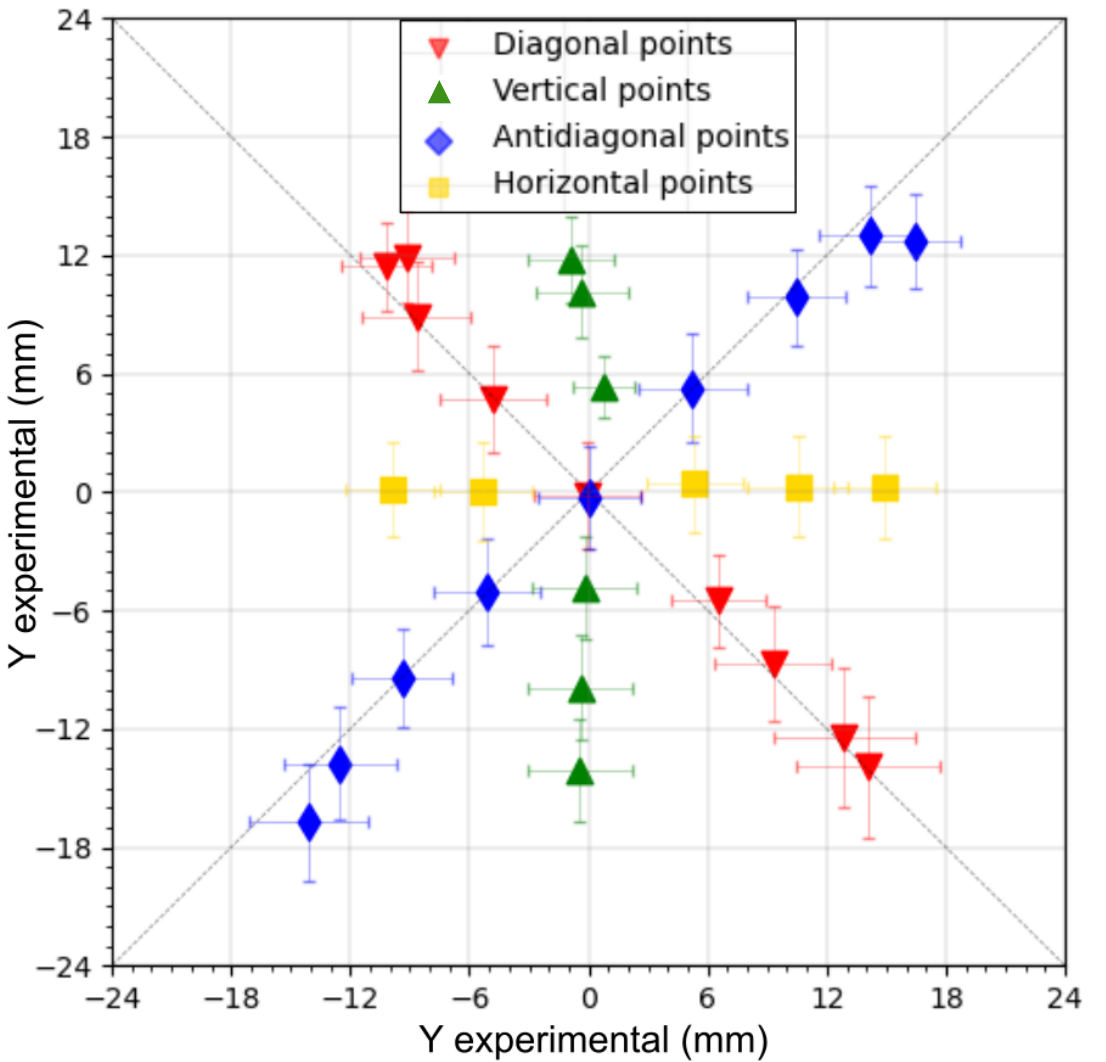}
\caption{Centroid of the distributions of reconstructed positions and standard deviation(error bar). Each point corresponds to one irradiation position and the different symbols represent the sets of of measurements along the vertical, horizontal and diagonals.}
\label{fig:Centroids}
\end{figure}

Fig.~\ref{fig:Centroids} depicts the centroid of the reconstructed positions for all the irradiation points measured within the CLYC crystal. Different colours and symbols have been used to distinguish various regions of the crystal. The error bars in Fig.~\ref{fig:Centroids} represent the standard deviation of the intrinsic spatial resolution extracted with the aforementioned methodology. The values of spatial resolution extracted from this characterization, typically expressed in terms of the FWHM, range from 4.5 to 6~mm, with most values around 5~mm (see the supplementary material). This result demonstrates that a sub-pixel spatial resolution has been achieved using a CLYC monolithic crystal coupled to a SiPM and the AIT 4-channel active base readout.

\begin{figure}[!t]
\centering
\includegraphics[width=0.8\linewidth]{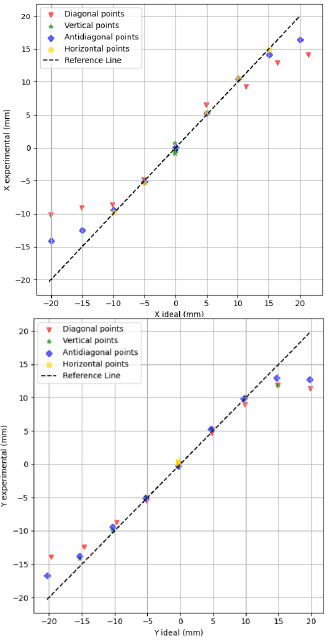}
\caption{Centroid of the reconstructed distributions as a function of the true coordinate. The top and bottom panel correspond, respectively, to the$\times$and y coordinate and different colors represent the same measurement subsets of Fig.~\ref{fig:Centroids}.}
\label{fig:Linearity}
\end{figure}

To complete the precise spatial characterization of the CLYC-SiPM detector, we studied the linearity of the response aiming to establish the effective surface in which positions can be reconstructed. The linearity diagrams, shown in Fig.~\ref{fig:Linearity}, are defined as the relationship between the mean value of the reconstructed coordinates for the individual irradiation points ($x/y_{experimental}$) and the true coordinates ($x/y_{ideal}$), which are delivered by the $xy$-gantry of the $\gamma$-ray scanning table. Fig.~\ref{fig:Linearity} shows in different colors the same four subsets of Fig.~\ref{fig:Centroids} associated to the vertical and horizontal axes and the two crystal diagonals. 

\begin{figure}[!htbp]
\centering
\includegraphics[width=0.85\linewidth]{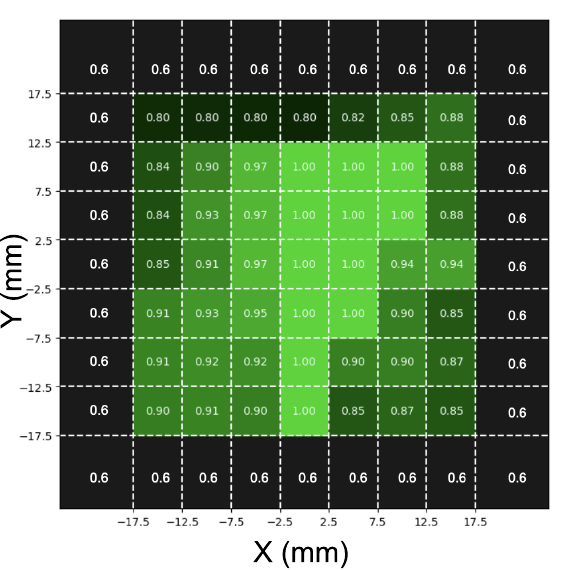}
\caption{Compression factor, defined as the ratio between the reconstructed coordinate and the true one as a function of the true irradiation. Only the Y coordinate is shown for simplicity.}
\label{fig:Compression}
\end{figure}

\begin{figure*}[!h]
\centering
\includegraphics[width=0.99\linewidth]{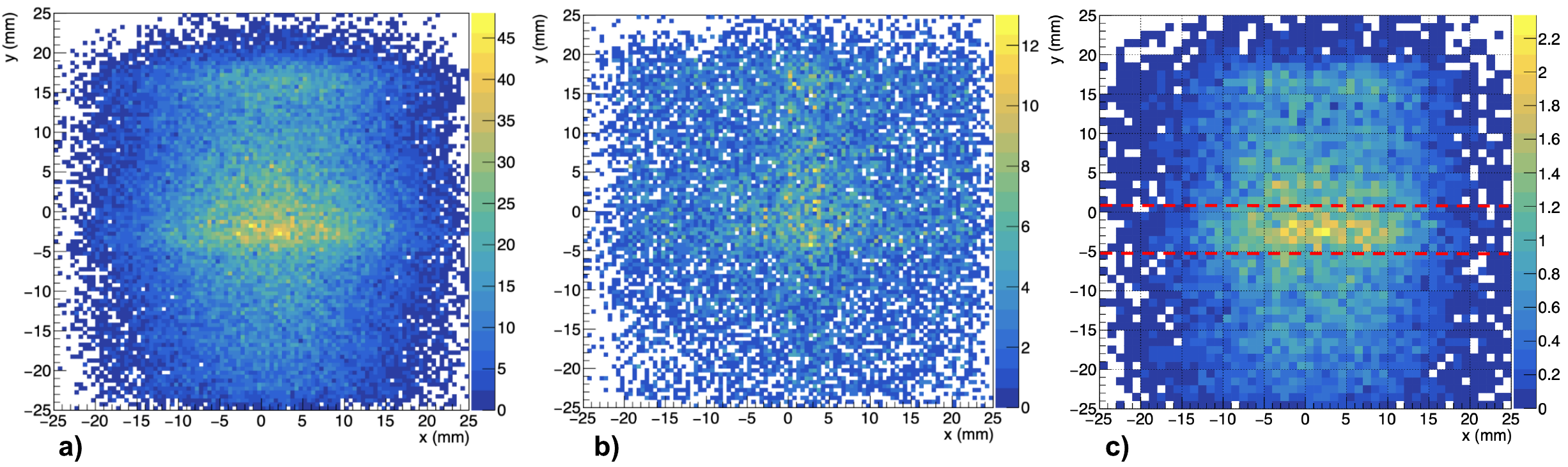}
\caption{Reconstructed positions of neutron hits during the irradiation with the partially moderated $^{252}$Cf source behind the slit pattern. Raw reconstructed pattern (a), background pattern (b) and background-subtracted result with the slit dimensions superimposed (c). }
\label{fig:NeutronPattern}
\end{figure*}

The results of Fig.~\ref{fig:Linearity} show that linearity is maintained at the center of the crystal. However, as the reconstructed position moves further from the center -- beyond 15 mm -- the compression effect becomes more significant, and no hits are reconstructed in the peripheral region. This result does not imply that the crystal is insensitive in this region; rather, the reconstructed positions are shifted towards the center. This effect, which has been discussed in detail in previous works~\cite{Babiano19,Balibrea:21}, is rather prominent for scintillation crystals such as the CLYC used here, where the walls are covered with a reflector. For such detectors the major contribution to the optical light distribution registered by the SiPM near the edges is due to reflections in the walls. Better linearity has been reported with the treatments using black paint, although this comes at the cost of a significant worsening of the energy resolution~\cite{Freire:20}. Alternative solutions in monolithic crystals with reflective layers require the implementation of analytical or Machine-Learning based corrections (see, for instance, Ref.~\cite{Balibrea:21}).

To visualize the magnitude of the compression effect across the crystal surface, in Fig.~\ref{fig:Compression} we represent the compression factor, defined as the ratio between the reconstructed coordinate and the true one. For positions where no measurements were performed, the values have been linearly interpolated. This diagram indicates that compression factors between 0.80  and 1 (i.e. no compression) are found in the central 30$\times$30 mm$^2$ of the crystal. The peripheral region, shown in black in this diagram, presents compression factors of around 0.60.

 The successful validation of the spatial sensitivity presented in this section represents the first detailed characterization of the attainable resolution with a CLYC crystal coupled to a SiPM array, although position sensitive CLYC detectors using PSPMTs have been previously developed~\cite{Soundara:12,Whitney:15}. When compared to previous works on monolithic inorganic detectors (LaCl$_3$, LaBr$_3$, LSO, etc.), the accuracy depends significantly on the pixel size and the methodology used to reconstruct the event~\cite{Li10,Pani:11,Lerche:05}. The closest approach to our multiplexing method is the resistor networks, for which values of 3.4~mm have been reported for a 42$\times$42~mm$^2$ LSO block coupled to a 8$\times$8 pixelated readout~\cite{Lerche:09}.
 Better resolutions in the transverse plane of 1-2~mm have been reported for individual pixel readout~\cite{Vandam:11,Babiano19,Balibrea:21} using analytical algorithms~\cite{Li10} and 2~mm resolutions have been obtained using 3-mm-pitch SiPM arrays coupled to a column-row readout (24 channels)~\cite{Freire:20}. 

\subsection{Reconstruction of neutron spatial patterns}\label{sec:Pos_NeutGamma}

The final aim of the CLYC detector in GN-Vision is to act as a position-sensitive detector with neutron-gamma discrimination capability. In the previous section, we have presented the characterization of the position sensitivity using a collimated $\gamma$-ray beam. In this section, we discuss the first tests position-sensitive neutron-gamma discrimination with our CLYC-SiPM detector leading to the construction of spatial patterns of neutron interactions.

The results presented hereafter were obtained with the experimental setup displayed in the bottom panel of Fig.~\ref{fig:Setup_Pos}. In this setup, thermal and epithermal neutrons emitted from the neutron moderator are expected to be absorbed in the $^6$LiPE slab, with the exception of those that travel through the slit and reach the position-sensitive CLYC. Thus, the reconstructed distribution of neutron hits in the detector is expected to show an interaction pattern that resembles the slit. In order to assess the possible background of scattered neutrons, an ancillary measurement was carried out with the slit closed. 

In the analysis of these measurements, both fast neutrons and $\gamma$-rays were filtered out with a double selection in PSD range and signal amplitude (i.e. deposited energy). The latter is particularly relevant to filter out fast neutrons emitted by $^{252}$Cf which are not thermalized in the moderator and travel through the $^{6}$LiPE absorbing material, increasing the background in the reconstructed neutron pattern. The applied cut around the slow neutron bump in the PSD-energy histogram is illustrated with red dashed lines in the insert of Fig.~\ref{fig:PSD_ADC_SiPM}.

Once the slow neutron events are selected, we reconstructed their interaction positions in the CLYC crystal on an event-by-event basis using the methodology applied for $\gamma$-rays, based on four position encoded signals (see Sec.\ref{sec:Pos_resol}). Fig.~\ref{fig:NeutronPattern} shows the reconstructed distributions of neutron hits for the measurement with the open slit (a) and the closed slit (b). The fact that the counts are concentrated in the central region for these two figures is explained from the proximity of the source and the detector. The third panel (c) shows the background-subtracted distribution. This figure clearly shows a pattern which is spatially correlated with the position of the slit in the neutron absorbing layer in front of the detector. The projection of the reconstructed pattern on the Y axis indicates that the attained signal-to-background ratio is around 3. 

The successful results presented in this section are the first experimental validation of the capability to carry out position-sensitive measurements of slow neutrons with the CLYC detector of GN-Vision. 

 \section{Proof-of-concept of neutron imaging }\label{sec:NeutronImaging}

 The proof-of-concept experiments of neutron imaging discussed herein were part of a larger experimental campaign carried out in June of 2024 at the FIPPS instrument~\cite{Michelagnoli:18} aimed to conduct a first pilot study of the applicability of Compton (gamma) and neutron imaging for dosimetry in BNCT~\cite{Lerendegui:24_AppRadIsot,Torres:24}. In this work, we present a portion of the measurements aimed at demonstrating the neutron-imaging capability and studying the attainable resolution. The experiment was performed in a thermal neutron beam since, according to the design phase of the device~\cite{Lerendegui:24}, this energy provides the best contrast and resolution in the obtained images. 
 
\subsection{Experiment at ILL}\label{sec:ExpILL}

The experimental setup for the neutron imaging validation was located in the experimental hall of FIPPS, which exploits the thermal neutron beam of the H22 guide exiting the research reactor of Institut Laue-Langevin (ILL). This reactor produces the most intense continuous neutron flux in the world in the moderator region, at 1.5$\cdot$10$^{15}$n/cm$^{2}$/s, with a thermal power of 58.3 MW. The thermal neutron beam of the H22 guide is collimated in vacuum up to the target position, located at the centre of the FIPPS array. The complex collimation system, which aims to reduce the beam-related $\gamma$-ray background in the detector, is composed of B$_4$C apertures followed by 5~cm lead absorbers and enriched $^6$LiF apertures. The latter are placed in the closest position to the FIPPS detectors~\cite{Michelagnoli:18}. The neutron beam at the target position has a diameter of 1.5~cm and a flux of 5$\cdot$10$^{7}$n/cm$^{2}$/s~\cite{Michelagnoli:18}.

\begin{figure}[!b]
\centering
\includegraphics[width=1.0\linewidth]{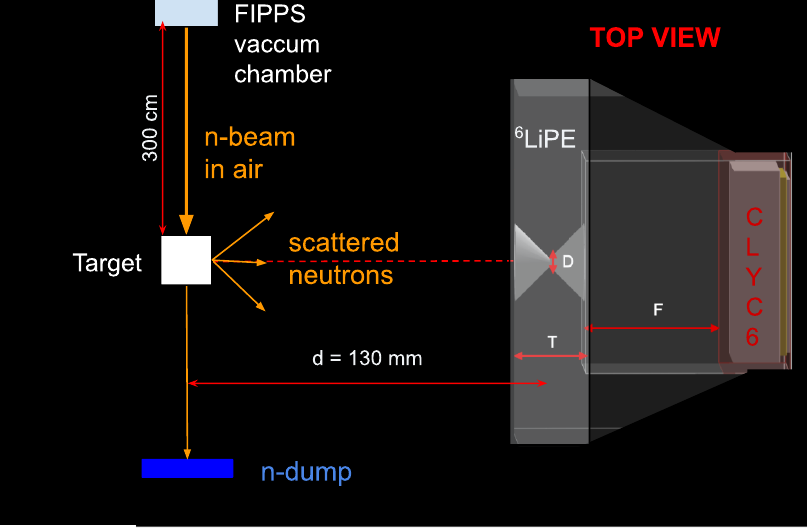}
\includegraphics[width=1.0\linewidth]{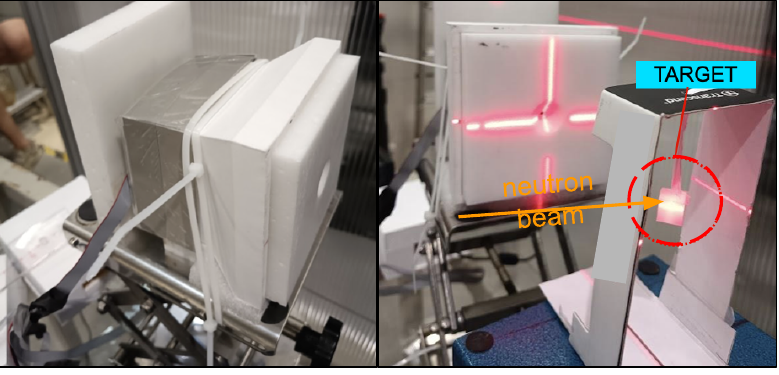}
\caption{Sketch (top) and picture (bottom) of the experimental setup at the H22 guide of ILL. The neutron imaging module (bottom left) was placed in-front of the object that were inserted in the neutron beam and the pinhole was vertically aligned with the object and beam (bottom right).}
\label{fig:ILLSetup}
\end{figure}

The neutron imaging setup was installed, as sketched in Fig.~\ref{fig:ILLSetup}, at about 3~m from the exit of the FIPPS vacuum chamber and at only 20~cm from the neutron dump, consisting of a concrete wall covered with a borated gum. The neutron imaging prototype used for this POC experiment, shown in the bottom panel of Fig.~\ref{fig:ILLSetup}, follows the initial design proposed for the implementation of the GN-Vision concept~\cite{Lerendegui:24}. The same CLYC-SiPM detector characterized in previous sections of this work was attached to a simple neutron pinhole collimator. For the absorbing material of the neutron collimator, we chose, among other low-Z neutron-absorbing materials, highly (95\%) $^{6}$Li-enriched polyethylene ($^6$LiPE) due to its high absorbing power and simple machining. Moreover, $^{6}$Li was chosen among the isotopes with large neutron absorption cross sections since no \g-rays are emitted in the $^{6}$Li(n,$\alpha$) reaction, thus avoiding a background source for the Compton imaging in the final GN-Vision device. The critical parameters of the neutron imaging device, are the thickness (T), diameter (D), and focal distance (F) of the pinhole. In the current prototype,  T = 20 mm, D = 2.5 mm and F = 40~mm. The reader is referred to the previous MC-based study~\cite{Lerendegui:24} for the details on the impact of these parameters in the imaging performance. Additional layers of $^{6}$LiPE, $^{6}$LiF and cadmium were placed surrounding the CLYC crystal and the space between the CLYC and the pinhole to reduce the background of scattered thermal neutrons, hence maximizing sensitivity to those coming through the pinhole.

The settings of the AIT 4-channel active base and the CAEN DAQ were the same as those used for the results in Sec.~\ref{sec:Pos_NeutGamma}. This consistency facilitated the application of the same methodology and cuts to select slow neutron events and reconstruct their interaction positions.

\subsection{Measurements and neutron images}\label{sec:NeutronImages}

The aim of this POC experiment was to place several point-like sources of thermal neutrons in front of the neutron imager and study whether their positions could be spatially localized. For this purpose, several H-rich targets, featuring a large neutron scattering cross section, were placed in a plane located at a distance d=130~mm from the center of the pinhole collimator and vertically aligned with it (see Fig.~\ref{fig:ILLSetup}). To study the imaging performance along one direction, these targets were moved in the field of view of the imaging device along the horizontal (neutron beam) axis. Three different sets of measurements were carried out with the following goals:

\begin{enumerate}
     \item  \textbf{Spatial localization}: A 1x1x1 cm\(^3\) PE cube was placed at different positions along the beam axis (from -2~cm to 2 cm in steps of 1~cm) with the aim of studying the sensitivity to the position of a single point-like source of thermal neutrons.

     \item  \textbf{Source dimension}: A 2x2x2 cm\(^3\) PE cube was placed at different positions along the beam with the aim of investigating how the reconstructed images scale with the dimensions of the neutron source and how does the image contrast vary.

     \item \textbf{Resolving power}: Last, the resolving power of the neutron imager was evaluated by irradiating two small rubber cubes of $\simeq$4 mm side lengths placed along the beam at 2 and 3 cm from each other.
 \end{enumerate}

\begin{figure}[!htbp]
\centering
\includegraphics[width=0.8\linewidth]{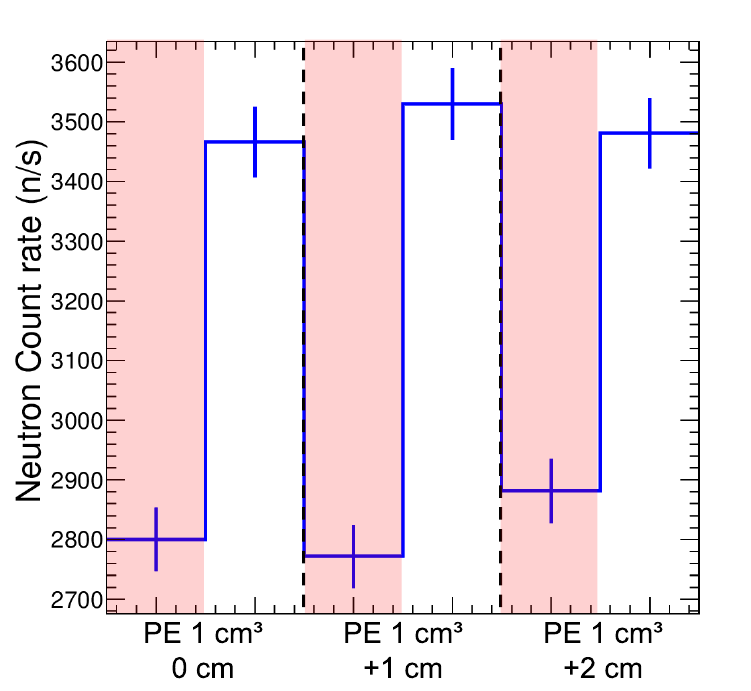}
\caption{Neutron count rate registered in three different irradiations of the 1x1x1 cm\(^3\) PE cube. The areas with the red shadow correspond to the background irradiations (i.e. pinhole blocked). }
\label{fig:NeutRate}
\end{figure}
 
     For each of the targets and positions we took measurements of only 60~s. In order to quantify the background neutrons (i.e. not coming through the pinhole) and assess our signal-to-background ratio we carried out a second measurement in each configuration with the pinhole blocked with an additional layer of $^{6}$LiPE. The data reduction process and neutron event selection methodology was the same one applied for the obtention of the first neutron patterns in Sec.~\ref{sec:Pos_NeutGamma}. Before extracting the reconstructed positions, we evaluated our sensitivity by studying the impact of opening the pinhole aperture in the registered neutron count rate (i.e. events selected after the PSD-amplitude cut). The results, shown in Fig.~\ref{fig:NeutRate} for irradiations of the 1~cm$^{3}$ PE cube in different positions, the neutron counts are enhanced in 25\% when the pinhole is opened, thus ensuring the contrast in the image. This has been possible only after the reduction of the neutron background events in a factor $\simeq$3 associated to the very careful shielding of the detector.
     
Once the slow-neutron events are selected and the 2D-coordinates of the neutron hits in the CLYC crystal are computed from the four encoded signals with Eq.~\ref{eq:codex}, the neutron image is reconstructed by simply applying inversions in both the X and Y axes, along with a scaling factor $S = d / F$. Here, $d$ is the distance from the collimator pinhole to the plane where the neutron sources are placed, and $F$ is the focal distance of the pinhole to the depth of interaction (DOI) of the neutron. Experimentally, the determination of the DOI is not possible with only four position signals; hence, we assume that neutrons interact in the front surface of the crystal, which is a reasonable premise for thermal neutrons in a CLYC. 

\begin{figure}[!t]
\centering
\includegraphics[width=0.85\linewidth]{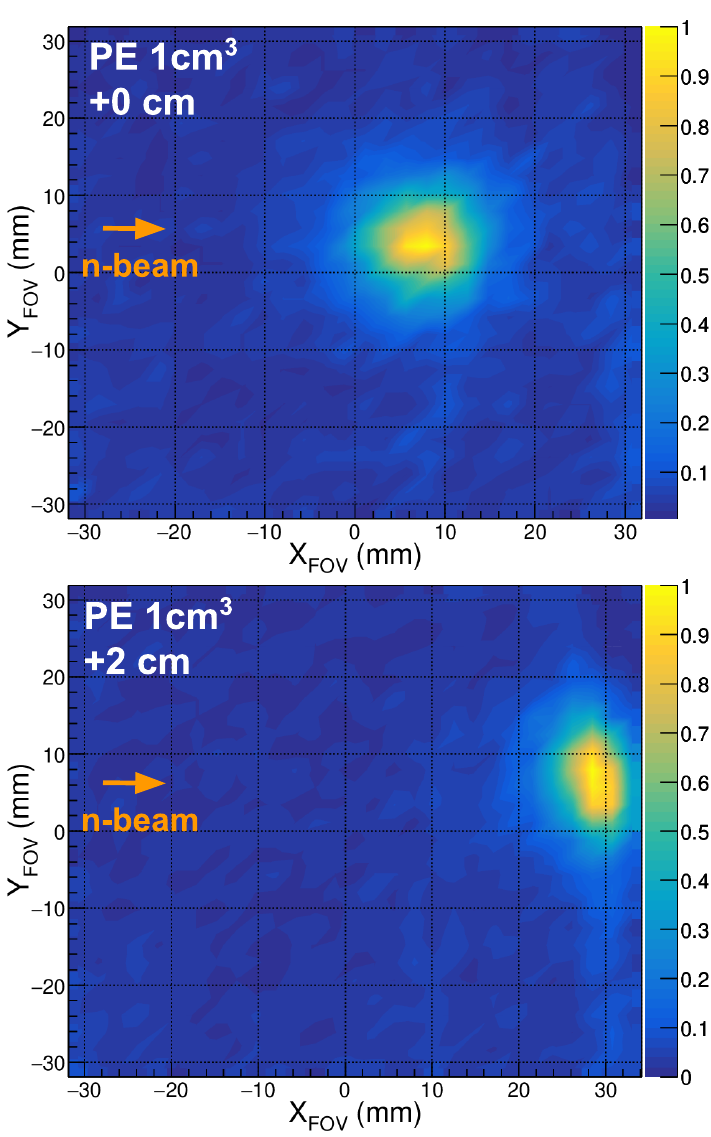}
\includegraphics[width=0.85\linewidth]{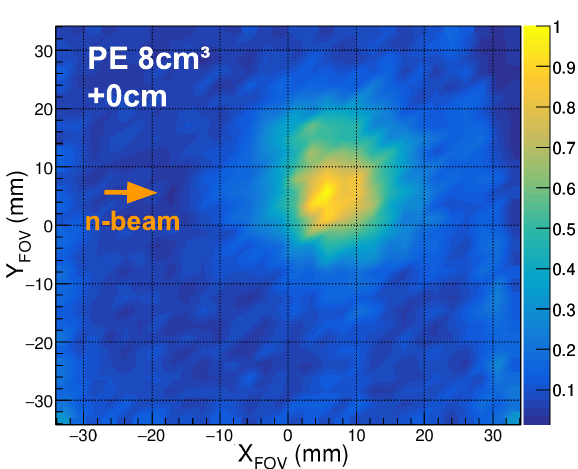}
\caption{Images reconstructed from the thermal neutrons scattered in a polyethylene cube of 1cm$^{3}$ placed at -2~cm (top) and centered (0 cm, center) along the beam axis  and image reconstructed for a centered 8cm$^{3}$ PE cube (bottom). }
\label{fig:Images_PECube}
\end{figure}

The first block of images was reconstructed for the irradiations of a PE cube of 1~cm$^3$ volume placed at different positions along the horizontal (beam) axis. The top and middle panels of Fig.~\ref{fig:Images_PECube} show two examples of the reconstructed neutron images, corresponding to the PE cube roughly centered with respect to the detector and shifted 2~cm upstream. These results confirm the imaging capability of the device since we observe a clear pattern in the image that follows the expected shift. Moreover, the dimensions of the object in the central position (11~mm FWHM) correspond to those of the PE cube, while some distortion is observed in the image of the shifted cube due to a shadowed strip on the reconstructed patters, probably related to a lack of optical contact with the SiPM. Despite the fact that no background has been subtracted from these images, a remarkable contrast -- peak-to-background  $\approx$ 15 -- has been obtained, following the expectations of the MC simulations of the ideal device~\cite{Lerendegui:24,Lerendegui:22_ANPC}. The spatial pattern associated to background neutrons is mainly concentrated in the crystal edges, which fall outside of the field of view displayed in Fig.~\ref{fig:Images_PECube}.

To study the sensitivity to the dimension of the neutron source, a further measurement involved the irradiation of a larger target, a 2$\times$2$\times$2 cm\(^3\) PE cube. The reconstructed image for the cube centered at the pinhole collimator is shown in the bottom panel of Fig~\ref{fig:Images_PECube}. By comparing the FWHM of the spots in the neutron images reconstructed for the 1 cm\(^3\) and 8 cm\(^3\) PE cubes, we observe that the size of the neutron image pattern on the detector is a factor 1.70 larger in the Y axis than that of the cube of 1~cm~edge. On the X axis (neutron beam axis), the scaling factor is only 1.48. This asymmetry may be related to the absorption of neutrons along the beam axis in the largest cube. Another sizable effect is the reduction in contrast, which decreases by a factor of approximately 2. This can be attributed to the larger total number of scattered neutrons, which increases the overall neutron background, and the partial absorption of those emerging from the cube in the direction of the collimator. 

\begin{figure}[!t]
\centering
\includegraphics[width=0.85\linewidth]{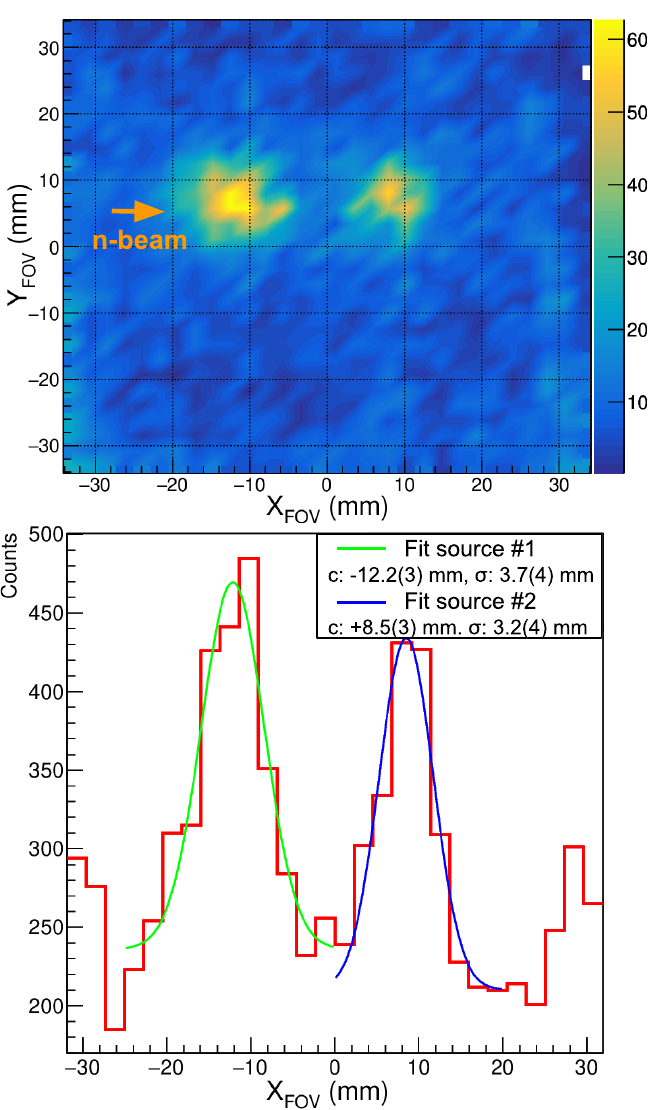}

\caption{Image reconstructed from the thermal neutrons scattered in two rubber cubes of 4~mm diameter aligned with the beam (top) and projections of the image onto the$\times$axis fitted to Gaussian profiles to extract the spatial resolution (bottom).}
\label{fig:NeutronImage_resolution}
\end{figure}

A last measurement in this POC experiment aimed at studying the resolving power by irradiating two small cubes of rubber placed along the beam axis a few centimeters apart. Fig.~\ref{fig:NeutronImage_resolution} shows the reconstructed 2D neutron image with the two cubes separated by only 2~cm. From this image it is clear than the cube located upstream scatters more neutrons. This result is also consistent with the fact that the downstream cube is expected to be significantly shadowed by the upstream one. The projections of the latter onto the beam axis, shown in the bottom panel of the same figure, indicate that the attainable resolution (FWHM) is $\sim$8.2~mm, which corresponds to to an angular resolution of $\sim$3.7$^{\circ}$. The ability to resolve two point-like neutron sources separated only 2~cm with GN-Vision, aligns with the expected performance from the MC simulations of the conceptual device~\cite{Lerendegui:22_ANPC,Lerendegui:24}. 

In terms of neutron image resolution, the results of these initial pilot experiments are comparable or better than the best neutron image resolutions (3-7$^{\circ}$) reported before for previous compact devices with dual neutron-gamma imaging capabilities~\cite{Boo:21,Guo:21}. The latter were mostly based on multiple pixels or layers of organic scintillators with the aim of imaging fast neutrons. Moreover, our resolving power seems quite promising when compared to the 9$^\circ$-30$^{\circ}$ resolutions reported for other compact dual imaging systems sensitive to fast neutrons~\cite{Steinberger:20} and for large scintillator arrays~\cite{Poitrasson:14,Poitrasson:15}. Focusing in devices where thermal and epithermal neutrons are detected, the results of the CLYC-based modification of a comercial gamma camera~\cite{Soundara:12,Whitney:15} look promising. However, no quantitative analysis of the resolution is reported in the latter. Finally, the multi-layer device of Ref.~\cite{Hamrashdi:20} is able to detect thermal neutrons but lacks of imaging capability. Last, recently deployed devices aimed only at neutron imaging-- mostly of fast neutrons-- also report similar resolutions~\cite{Liu:24,Lopez:22,Lynde:20} to those obtained here.

\section{Summary and outlook}\label{sec:Summary}

GN-Vision is an innovative device that combines neutron and gamma imaging capability in a single, compact and lightweight device. These properties make it attractive for medical applications, such as proton range and neutron dose monitoring in hadron therapy~\cite{Lerendegui:22_ANPC}. Moreover, the imaging of neutrons and \g-rays is also of great interest for nuclear safety and control and for nuclear security inspections, where sensible materials naturally emit both neutrons and \g-rays~\cite{Lerendegui:24}. The proposed device consists of two planes of position-sensitive detectors, based on monolithic LaCl$_{3}$ and CLYC-6 crystals, which exploits the Compton technique for \g-ray imaging. A mechanical lightweight collimator attached to the first plane enables the imaging of slow neutrons ($<$~100 eV). 

The first detection plane, comprising a CLYC-6 detector read out with an array of SiPMs to achieve position sensitivity, is the key element of this device towards the dual imaging capability. In this work, we have presented the first experimental milestones in the development of the CLYC-SiPM detector of GN-Vision. The energy resolution and performance in terms of PSD between neutrons and $\gamma$-rays was first evaluated for two different monolithic CLYC-6 crystals of 50$\times$50~mm$^2$ area and thicknesses of 8 mm (CLYC-A) and 13~mm (CLYC-B). The latter yielded the best resolution of 6.2\% when coupled to a PMT that degraded to 8.9\% when the crystal was attached to an 8$\times$8 array SiPM and the sum signal of the 64 pixels was used to perform spectroscopy. As for the PSD, the CLYC-B led also to the best separation between thermal neutrons and $\gamma$-rays, characterized by FOM values of 3.8 and 2.9, respectively, with a PMT and SiPM readout. The characterization of the spatial response of the CLYC-SiPM assembly has yielded a sub-pixel position resolution, with a most probable value of 5~mm, for an array of 8$\times$8 SiPMs of 6~mm pitch. Moreover, a linear response has been found in the central 30~mm. Last, we have studied also the spatial sensitivity to slow neutrons, leading to the first reconstructed neutron patterns, thanks to the accurate removal of the gamma and fast neutron background.

The development and characterization of the position-
sensitive neutron-gamma discriminating CLYC detector, has built the foundation for the first proof-of-concept experiment of the neutron imaging capability of GN-Vision. For such experiment, the CLYC-SiPM detector was assembled to a $^6$LiPE neutron pinhole collimator. The first POC experiments were
carried out at ILL-Grenoble using the scattering of a thermal neutron beam in small plastic targets. A remarkable angular resolution of 4$^{\circ}$ and a peak-to-background of almost a factor of 20 have been obtained in this neutron imaging tests utilizing a beam of thermal neutrons. The
successful results of these experiments, which aligned with
the expected performance from MC simulations, represent the
first major experimental breakthrough in the development of the
GN-Vision device.

In parallel to the experimental development discussed herein, several studies based on MC simulation are being carried out with the aim of improving the final design of GN-Vision and overcoming the main limitations of the current prototype, particularly those related to the collimation system. In this context, we are studying the possibility of enhancing the efficiency in at least one order of magnitude by using a coded-aperture mask~\cite{JamesThesis}. In order to make GN-Vision a more cost-effective solution, we are also studying the impact of replacing the 95\% enriched $^{6}$LiPE with natural LiPE for various applications~\cite{AndreaThesis}. The following steps in the development will cope with the integration of the dual imaging technique. While the $\gamma$-imaging capability is already at a very high technology
readiness level (TRL) of 6 following the developments of the previous i-TED Compton imager and its use in medical physics applications~\cite{Lerendegui:22,Lerendegui:22_ANPC,Balibrea:22,Lerendegui:24_AppRadIsot,Torres:24}, the neutron imaging capability has just been experimentally proven for the first time in this work. The combination of both imaging modalities will require the integration of the compact PETSys TOFPET2 electronics, used to readout the LaCl$_3$ crystals of the absorber plane, with the output of the AIT readout circuit of the position-sensitive CLYC. Lastly, more realistic test measurements in proton therapy and BNCT facilities as well as in relevant scenarios for inspections of SNM will follow in the upcoming years.

\section*{Declaration of competing interest}
The authors declare that they have no known competing financial interests or personal relationships that could have appeared to influence the work reported in this paper.
% https://www.elsevier.com/authors/journal-authors/policies-and-ethics/credit-author-statement

\section*{CRediT authorship contribution statement}
\textbf{J. Lerendegui-Marco:} Conceptualization, Investigation, Methodology, Supervision, Formal analysis, Data curation, Visualization, Writing - original draft, Project administration, Funding acquisition.
\textbf{G. Cisterna:} Investigation, Data curation, Visualization.
\textbf{J. Hallam:} Investigation, Data curation, Visualization, Writing -review \& editing.
\textbf{V.~Babiano:} Investigation, Software, Writing -review \& editing. 
\textbf{J. Balibrea-Correa:} Investigation, Formal analysis, Software.
\textbf{D.~Calvo:} Investigation, resources.
\textbf{G.~de la Fuente:} Investigation. 
\textbf{B. Gameiro:} Investigation, Software. 
\textbf{I. Ladarescu:} Software. 
\textbf{A. Sanchis-Molt\'o:} Investigation, Writing -review \& editing. 
\textbf{P. Torres-S\'anchez:} Investigation. 
\textbf{C. Domingo-Pardo:} Conceptualization, Investigation, Methodology, Supervision, Writing -review \& editing, Project administration, Funding acquisition.

\section*{Acknowledgments}
This work builds upon research conducted under the ERC Consolidator Grant project HYMNS (grant agreement No. 681740) and has been supported by the ERC Proof-of-Concept Grant project GN-Vision (grant agreement No. 101113330). We acknowledge funding from the Universitat de Val\`encia through the Valoritza i Transfereix Programme, under grant UV-INV\_PROVAL21-1924580. We also acknowledge funding from the Spanish Ministerio de Ciencia e Innovación under grants PID2022-138297NB-C21 and PID2019-104714GB-C21, as well as from CSIC under grant CSIC-2023-AEP128. We acknowledge support from the Severo Ochoa Grant CEX2023-001292-S funded by MCIU/AEI.
Additionally, the authors thank the support provided by postdoctoral grants FJC2020-044688-I and ICJ220-045122-I, funded by MCIN/AEI/10.13039/501100011033 and the European Union NextGenerationEU/PRTR;  postdoctoral grant CIAPOS/2022/020 funded by the Generalitat Valenciana and the European Social Fund and a PhD grant PRE2023 from CSIC. Funding from the Severo Ochoa project under grant CEX2023-001292-S funded by MCIU/AEI, and financial support from the Institute Laue Langevin during the experimental campaign are also gratefully acknowledged.

%\bibliography{bibliography}

\end{document}